\documentclass[10pt,english,aps,prb,showkeys,superscriptaddress,floatfix,twocolumn]{revtex4-2}
\usepackage[dvipsnames]{xcolor}
\usepackage{color,xcolor}
\usepackage{placeins}
\usepackage{textcomp}
\usepackage{ae,aecompl}
\usepackage[T1]{fontenc}
\usepackage[utf8]{inputenc}
\usepackage{babel}
\usepackage{amsmath}
\usepackage{amssymb}
\usepackage{graphicx}
\usepackage{wasysym}
\usepackage{subcaption}
\usepackage[normalem]{ulem}
\usepackage{lipsum}
\usepackage[unicode=true,bookmarks=true,bookmarksnumbered=false,bookmarksopen=false,breaklinks=false,pdfborder={0 0 1},backref=false,colorlinks=true,linkcolor=magenta,citecolor=blue]{hyperref}
\usepackage{contour}
\DeclareMathAlphabet{\mathpzc}{OT1}{pzc}{m}{it}
\begin{document}
\global\long\def\ket#1{\left|#1\right\rangle }%
\global\long\def\bra#1{\left\langle #1\right|}%
\global\long\def\braket#1#2{\langle#1|#2\rangle}%
\global\long\def\expectation#1#2#3{\langle#1|#2|#3\rangle}%
\global\long\def\average#1{\langle#1\rangle}%
\title{Single impurity-induced localization transitions in electronic systems}
\author{Niaz Ali Khan}
\affiliation{Department of Physics, Xiamen University, Xiamen 361005, China}
\author{Munsif Jan}
\affiliation{ CAS Key Laboratory of Quantum Information, University of Science and Technology of China, Hefei 230026, China }
\affiliation{Hefei Unitary Quantum Technology Co., Ltd., Hefei 230088, China}
\author{Muzamil Shah}
\affiliation{Department of Physics, Quaid-i-Azam University Islamabad 45320, Pakistan}
\author{Muhammad Sajid}
\email{m.sajid@kust.edu.pk}
\affiliation{Department of Physics, Kohat University of Science and Technology, Kohat 26000, Khyber Pakhtunkhwa, Pakistan}
\author{Muhammad Mateen}
\affiliation{Department of Physics, Jangsu University, Zhenjiang 212013, China}
\author{Mushtaq Ali}
\affiliation{Department of Physics, Division of Science and Technology, University of Education Lahore, Pakistan}
\begin{abstract}
Anderson localization is a fundamental phenomenon in disordered quantum systems, where transport is suppressed by wave interference from extensive randomness. Moving beyond traditional multi-impurity scenarios, we investigate impurity-induced localization phenomena in low-dimensional tight-binding systems by focusing on the properties of impurity-generated bound states. By introducing a single on-site impurity into an otherwise extended lattice, we demonstrate that the impurity can host a bound state whose spatial character undergoes a transition from extended to localized as the impurity strength surpasses a critical value. This transition pertains solely to the impurity state, while the bulk states of the host system remain extended. We characterize the localization behavior by analyzing two distinct spatial profiles of the bound states: one with symmetric decay and another with exponential decay from the impurity site. Our results highlight how a local perturbation can induce nontrivial localization behavior at the level of individual eigenstates, without implying a global localization transition of the underlying electronic system.
\end{abstract}
\maketitle

\section{Introduction}

A quantum phase transition (QPT) is an abrupt change in the ground state of a many-body system driven solely by quantum fluctuations at zero temperature. Such transitions provide a fundamental framework for understanding diverse quantum phenomena across physical systems. Prominent examples include the superfluid-to-Mott-insulator transition in ultracold atoms \cite{greiner2002quantum}, which has become a paradigmatic platform for quantum simulation \cite{bloch2022superfluid}, as well as related studies in cavity optomagnonic systems \cite{cao2022}. QPTs also underpin critical behavior in spin systems, such as the ferromagnetic-to-paramagnetic transition \cite{PhysRevLett.126.116401,PhysRevB.81.205102}, and various manifestations of metal-insulator transitions \cite{paschen2021quantum,zheng2022quantum}.

Anderson localization is a fundamental phenomenon in which uncorrelated disorder causes electronic wavefunctions to become spatially confined rather than extended throughout a material \cite{Anderson1958,Thouless1979,Pires2019,Niaz2021,Economou2006}. This localization of eigenstates has profound consequences, most notably the suppression of transport, and has been observed across diverse platforms including disordered solids \cite{PhysRevB.95.144204,poduval2023anderson}, optical fibers \cite{Mafi2019}, and ultracold atoms in optical lattices \cite{white2020observation,jendrzejewski2012three}. In low-dimensional, non-interacting tight-binding systems, even weak disorder can strongly affect transport properties by inducing localization of single-particle states \cite{Thouless1979,Niaz2019,Niaz2020,AGNE20212970}. In contrast, three-dimensional systems may exhibit a mobility edge separating extended and localized eigenstates, requiring a finite critical disorder strength \cite{Niaz2019,Niaz2020,AGNE20212970,Roati2008,Muller2010,jendrzejewski2012three,Lagendijk2009,Niaz2022,Niaz2022-EC1DCD}. Beyond equilibrium properties, the interplay between disorder and quantum dynamics gives rise to rich non-equilibrium phenomena, including dynamical phase transitions manifested in the time evolution of quantum states and correlations \cite{Niaz2023,Ye2024,Niaz2024,KHAN2024114975,Ye2025}.

A closely related yet conceptually distinct problem concerns impurity-induced localization, where a local perturbation can qualitatively modify the properties of specific eigenstates without inducing a global localization transition of the host system. In this context, a single-impurity-induced quantum transition refers to a critical point at which an impurity-generated bound state undergoes a qualitative change in its spatial character as system parameters are varied. A paradigmatic framework for such studies is provided by impurity models, including the Anderson impurity model \cite{PhysRevB.107.125103,Roccati2021,Fabrizio2022,Davies2025,Molignini2023,Wang2025,Hetenyi2025,Leumer2025}. In these systems, a localized bound state may emerge when the impurity potential exceeds a critical strength, while the remaining bulk states of the lattice remain extended. Recent extensions of the single-impurity problem to non-Hermitian lattices, such as the nonreciprocal Su–Schrieffer–Heeger \cite{Liu2020} and Hatano–Nelson models \cite{Liu2021}, have revealed strong sensitivity to boundary conditions and size-dependent non-Hermitian skin effects. However, to the best of our knowledge, the role of a single impurity in inducing localization transitions of impurity-bound states in non-interacting Hermitian lattice systems under periodic boundary conditions has not been systematically explored.

In this work, we investigate the influence of a single on-site impurity potential on non-interacting lattice models with periodic boundary conditions. Our analysis focuses on the properties of impurity-generated eigenstates rather than on a global localization transition of the system. We characterize the impurity-induced states using the inverse participation ratio (IPR), together with the spatial structure of the corresponding eigenfunctions. We find that the impurity-site IPR scales as $\mathcal{O}(1/L)$ when the impurity-bound state is extended across the lattice, while it approaches unity in the regime where the state becomes strongly localized around the impurity, with $L$ denoting the system size. Correspondingly, the eigenstate profiles exhibit a clear crossover from uniform spatial distributions to exponentially localized behavior centered at the impurity site. For Hermitian systems, the localized impurity states display symmetric spatial profiles consistent with the underlying lattice symmetry.

The structure of the paper is as follows. Section~\ref{sec:Model} introduces the tight-binding model with a single impurity and discusses the inverse participation ratio as a diagnostic tool for characterizing eigenstate localization. Section~\ref{sec:ImpurityInducedLocalization} presents a detailed analysis of impurity-induced localization transitions of bound states under periodic boundary conditions. Finally, we summarize our conclusions and outlook in the last section.

\section{The Model\label{sec:Model}}
The model we focus on consists of noninteracting spinless electrons in the presence of a single-site impurity under periodic boundary conditions (PBC). The periodic geometry eliminates boundary-induced localization effects, allowing to isolate the role of the impurity in modifying the properties of individual eigenstates. However, this choice does not alter the qualitative impurity-induced localization behavior. The physical interpretation of the results is therefore robust against the choice of boundary conditions. The Hamiltonian in one-dimension (1D) lattice has the general form \citep{Liu2020,Liu2021},
\begin{equation}
\hat{\mathcal{H}}=t\sum_{n=0}^{L}(c_{n}^{\dagger}c_{n+1}+c_{n+1}^{\dagger}c_{n})+\varepsilon_{0} c_{0}^{\dagger}c_{0},\label{eq:1DHamiltonian}
\end{equation}
where $\varepsilon_{0}$ denotes the amplitude of the single site energy of an electron at $0$th site of the lattice of size $L$ and $t$ is the transfer energy (hopping integrals) between the nearest neighboring
sites. The transfer integrals are set to unity $t=1,$ and all energy scales are measured in units of $t$. In order to characterize the localized or extended nature of the impurity-induced bound state (normalized) $\ket{\psi_{i,n}}$, we calculate its inverse participation ratio ($\text{IPR}_{n}$),
\begin{align}
\text{IPR}_{n} = \sum_{i=1}^{L} \left|\Psi_{i,n}\right|^{4}.\label{eq:IPR-General}
\end{align}
The $\text{IPR}_{n}$ reveals the transport characteristics of $\ket{\psi_{i,n}}$. Its system-size scaling follows distinct power laws in different phases:
\begin{figure}[ht!]
\begin{centering}
\includegraphics[scale=0.36]{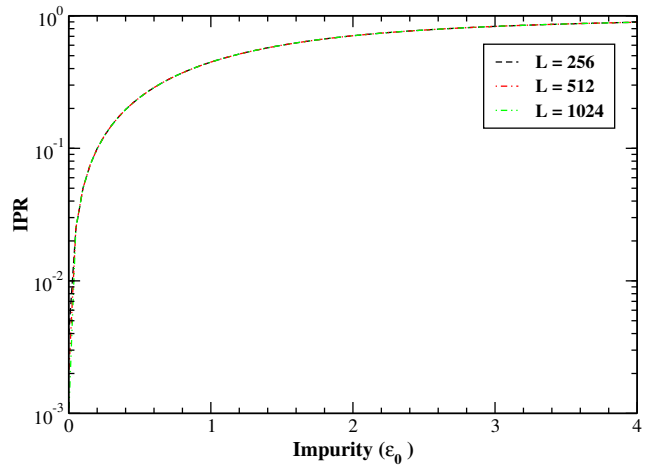}
\par\end{centering}
\caption{(Color online) The IPR of the bound state as a function of the impurity strength $\varepsilon_{0}$ of the 1D single-impurity model. In the presence of an infinitesimal impurity strength, the eigenstate is localized, corresponding to the size-independent IPR $\approx 1$.\label{fig:1D-ipr}}
\end{figure}
\begin{figure}[ht!]
\begin{centering}
\includegraphics[scale=0.36]{1D-State.eps}
\par\end{centering}
\caption{(Color online) Log-linear scale:} The impurity-induced bound eigenstate distributions of the 1D model with various strengths of impurity amplitude. The skewness of the distributions is zero, corresponding to symmetrical eigenstate distributions.\label{fig:1D-State}
\end{figure}
\begin{itemize}
    \item Extended phase: $\text{IPR}_{n} \propto L^{\alpha}$ with $\alpha \approx -1$, reflecting the delocalized nature of wavefunction, i.e. it extends over the entire lattice. The IPR tends to $0$ in the thermodynamic limit in this phase. 
    \item Localized phase: $\text{IPR}_{n} \propto  L^{\alpha}$ with $\alpha=0$, which is a characteristic of wavefunctions exponentially confined to a finite region. In the localized phase, the IPR tends to $1$ in the thermodynamic limit.
    \item Critical phase: $\text{IPR}_{n} \propto  L^{\alpha}$ with $\alpha\approx0$. The IPR in this phase is smaller than $1$ but is much higher than zero. The wavefunction is neither fully extended nor exponentially localized.
\end{itemize}
The exponent $\alpha$ thus serves as a scaling dimension that distinguishes between metallic, insulating, and critical regimes.
\section{The Impurity-induced Localization\label{sec:ImpurityInducedLocalization}}
\begin{figure}
\begin{centering}
\includegraphics[scale=0.38]{2D-ipr.eps}
\par\end{centering}
\caption{(Color online) (a) The IPR (log-linear scale) of the bound state as a function of the impurity strength $\varepsilon_{0}$ in log-linear scale of the 2D single-impurity model. The nature of the bound state changes from extended to localized by tuning the impurity strength. In the inset, we enlarge the IPR region around $\varepsilon_0=1-2$ to clarify the finite-size critical regime. (b) The IPR (log-log scale) of the bound state versus system size $L$ for fixed impurity strength. The data is very well fitted by $\text{IPR}\propto L^{\alpha}$. The IPR is approximately equal to $1$ and is independent of $L$ for $\varepsilon_0=3.0$ ($\alpha=0$), varies algebraically with $L$ for $\varepsilon_0=1.5$ ($\alpha\approx0$), and decays for $\varepsilon_0=0.5$ ($\alpha<0$), corresponding to localized, critical, and extended behavior of the state, respectively.\label{fig:2D-ipr}}
\end{figure}
\begin{figure}
\begin{subfigure}[h]{0.96\linewidth}
\includegraphics[width=\linewidth]{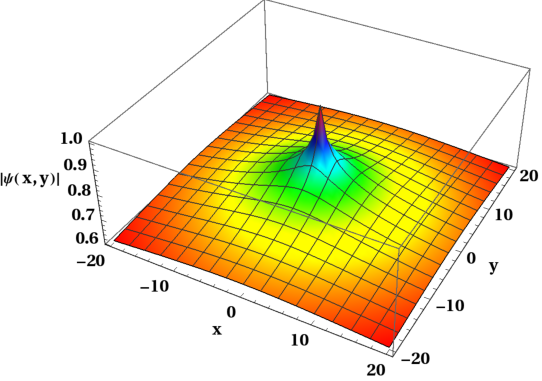}
\caption{$\varepsilon_{0} = 0.5$ (Extended regime)}
\end{subfigure}
\hfill
\begin{subfigure}[h]{0.96\linewidth}
\includegraphics[width=\linewidth]{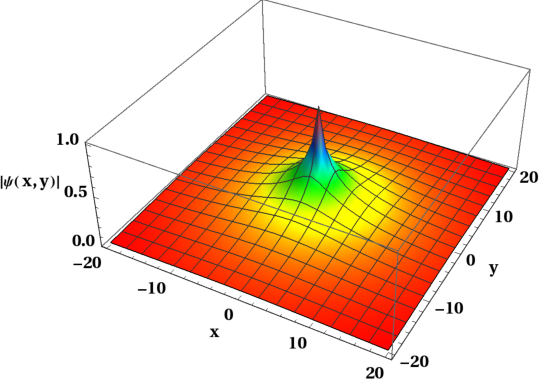}
\caption{$\varepsilon_{0} = 1.5$ (Critical regime)}
\end{subfigure}%
\hfill
\begin{subfigure}[h]{0.96\linewidth}
\includegraphics[width=\linewidth]{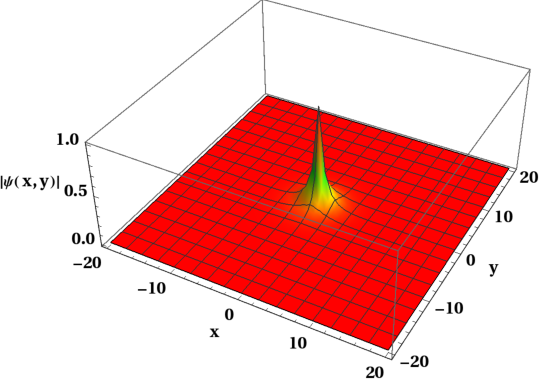}
\caption{$\varepsilon_{0} = 3$ (Localized regime)}
\end{subfigure}%
\caption{The spatial distributions of the impurity-induced bound state of the} 2D impurity model with impurity amplitude (a) $\varepsilon_{0}=0.5$, (b) $\varepsilon_{0}=1.5$ and (c) $\varepsilon_{0}=3$. The distributions are symmetrical around (x,y)=(0,0).\label{fig:2D-Eigenstates}
\end{figure}
%
\subsection{The 1D Impurity Lattice}
The main focus is to study the single-impurity problem on a lattice
system under PBCs. In the absence of impurity potential $(\varepsilon_{0}=0),$ the eigenstates of the Hamiltonian $\mathcal{H}_{\text{1D}}$ are plane waves with eigenenergies, $E_{\text{1D}}=2t\cos k$.
Figure~\ref{fig:1D-ipr} shows the IPR of the bound state for various system sizes as a function of impurity amplitude. The typical value of the IPR monotonically increases with impurity strength as it approaches unity, characterizing the localized nature of the system. Moreover, the IPR turns out to be size-independent in the presence of infinitesimal impurity strength. This reflects the idea of a single-site localization transition at zero impurity strength in a 1D lattice. In the presence of the single-impurity potential, the eigenstate of the Hamiltonian can be obtained through the Green\textquoteright s functions:
\begin{equation}
\psi(x)=\frac{\varepsilon_{0}\delta_{x,0}}{E_{\text{1D}}-\mathcal{H}_{\text{1D}}}\psi(0),\label{eq:1D-GreenFtn}
\end{equation}
Making a Fourier transformation of $\psi(x)$, we get,
\begin{align}
\psi(k) & =\frac{\varepsilon_{0}\psi(0)}{E_{\text{1D}}-2t\cos k},\label{eq:}
\end{align}
Transforming back to the position space by making the inverse Fourier
transformation of $\psi(k)$ gives,
\begin{align}
\psi(x) & =\frac{1}{2\pi}\intop_{-\pi}^{\pi}dk\frac{e^{ikx}\varepsilon_{0}\psi(0)}{E_{\text{1D}}-2t\cos k},\label{eq:-1}
\end{align}
The eigenenergy of the bound state can be obtained at the impurity
embedded site $(x=0)$ as,
\begin{align}
E_{\text{1D}} & =\pm\sqrt{4t^{2}+\varepsilon_{0}^{2}},\label{eq1}
\end{align}
The corresponding eigenstates are
\begin{equation}
\psi(x)=\frac{\varepsilon_{0}\psi(0)}{E_{\text{1D}}-2t}\,{}_{3}\tilde{F}_{2}(\frac{1}{2},1,1;1-x,1+x;\frac{4t}{E_{\text{1D}}-2t}),
\end{equation}
where $_{3}\tilde{F}_{2}(z)$ is a regularized generalized hypergeometric
function. Figure~\ref{fig:1D-State}, demonstrates the distributions of eigenstates for the single-site embedded model in one dimension. The eigenstate distributions decay exponentially with distance from the location of the impurity. The decay becomes more profound as the strength of the impurity increases. Moreover, the eigenstate distributions are symmetrical around zero and can be characterized by calculating the mean distance, $\average{n}= \sum_{n}\braket{\psi}{n|\psi}$. The zero mean distance corresponds to the symmetrical distributions of the eigenstates. The eigenstate distributions also reveal the localized or extended nature of the quantum state at the corresponding site. One can obtain sharp, peaked distributions of eigenstates at impurity-enhanced sites for large impurity amplitudes. In ref. \citep{Liu2020}, the authors investigated the eigenstate distributions of a 1D non-Hermitian single-impurity lattice. It was found that positive or negative non-Hermiticity may trigger right-skewed distributions (positive skewness) or left-skewed distributions (negative skewness), respectively.
\begin{figure}[ht!]
\begin{centering}
\includegraphics[scale=0.36]{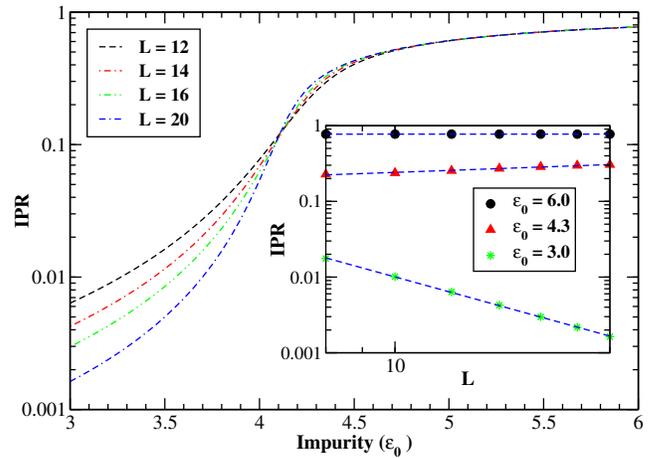}
\par\end{centering}
\caption{(Color online) The IPR of the impurity induced state as a function of the impurity strength $\varepsilon_{0}$ in 3D. The inset shows (on log-log scale) the IPR of the bound state as a function of linear system size $L$ at fixed impurity strength. A fit (dashed blue lines) $\text{IPR}\approx L^{\alpha}$ reveals distinct scaling exponent: for $\varepsilon_0=6.0$, $\alpha=0$ in localized regime; for $\varepsilon_0=1.5$, $\alpha\approx 0$ in the critical regime; and for 
$\varepsilon_0=0.5$, $\alpha<0$ in the extended regime.\label{fig:3D-ipr}}
\end{figure}
\subsection{The 2D Impurity Lattice}
This section is focused on the study of the single-impurity problem on a 2D lattice
system under PBCs. Without the impurity 
($\varepsilon_{0}{=}0$), the system system Hamiltonian $\mathcal{H}_{\text{2D}}$ is translationally invariant. Its eigenstates are therefore plane waves, and the corresponding energies follow the standard tight‑binding dispersion $E_{\text{2D}}=2t(\cos k_{x}+\cos k_{y})$. The eigenstates are translationally invariant, with probability amplitudes extended over all lattice sites, reflecting the metallic nature of the system. However, introducing single-impurity potential in the system results in a bound state, as illustrated in Fig.~\ref{fig:2D-ipr}. For $\varepsilon_{0}\le1.2$, the IPR of the impurity-induced state is size-dependent, and approaches zero as $L^{\alpha}$ with $\alpha<0$. On the other hand, the IPR scales as $L^{\alpha}$ in the critical regime ($1.2\lessapprox\varepsilon_{0}\lessapprox1.6$) with $\alpha\approx0$, and size-independent ($\alpha=0$) in the localized regime with $\varepsilon_{0}\gtrapprox1.6$. The eigenstate of the 2D Hamiltonian in the presence of the single-impurity potential can be obtained through the Green\textquoteright s
Functions:
\begin{equation}
\psi(x,y)=\frac{\varepsilon_{0}\delta_{x,0}\delta_{y,0}}{E_{\text{2D}}-\mathcal{H}_{\text{2D}}}\psi(0,0),\label{eq:2D-GreenFtn-1}
\end{equation}
The quasimomenta corresponding to the eigenstates $\psi(x,y)$ can
be obtained by making a Fourier transformation of $\psi(x,y)$,
\[
\psi(k_{x},k_{y})=\intop_{\Omega}dxdy\psi(x,y)e^{-i(k_{x}x+k_{y}y)},
\]
We get,
\begin{align}
\psi(k_{x},k_{y}) & =\frac{\varepsilon_{0}\psi(0,0)}{E_{\text{2D}}-2t(\cos k_{x}+\cos k_{y})},\label{eq:-2}
\end{align}
Transforming back to the position space by making the inverse Fourier transformation of $\psi(k_{x},k_{y})$ gives \cite{Kogan2024},
\begin{align}
\psi(x,y) & =\frac{1}{(2\pi)^{2}}\intop_{-\pi}^{\pi}\intop_{-\pi}^{\pi}dk_{x}dk_{y}\frac{\varepsilon_{0}\psi(0,0)e^{i(k_{x}x+k_{y}y)}}{E_{\text{2D}}-2t(\cos k_{x}+\cos k_{y})},\label{eq:-1-1}
\end{align}
At the impurity embedded site $(x=y=0)$, the expression~[Eq.~(\ref{eq:-1-1})] becomes
\begin{align}
\frac{\varepsilon_{0}}{(2\pi)^{2}}\intop_{-\pi}^{\pi}\intop_{-\pi}^{\pi}\frac{dk_{x}dk_{y}}{E_{\text{2D}}-4t\cos\frac{k_{x}+k_{y}}{2}\cos\frac{k_{x}-k_{y}}{2}} & =1,\label{eq:2}
\end{align}
Eq.~(\ref{eq:2}) can be expressed as
\begin{align}
\frac{\varepsilon_{0}}{2\pi^{2}}\intop_{-\pi}^{\pi}\intop_{0}^{\pi}\frac{d\alpha d\beta}{E_{\text{2D}}-4t\cos\alpha\cos\beta} & =1,\label{eq:2-1}
\end{align}
with
\begin{align*}
\alpha=\frac{k_{x}+k_{y}}{2},\qquad & \beta=\frac{k_{x}-k_{y}}{2},
\end{align*}
The expression~[Eq.~(\ref{eq:2-1})] have the following simplified
form
\begin{equation}
\frac{K\left(\frac{16t^{2}}{16t^{2}-E_{\text{2D}}^{2}}\right)}{\sqrt{E_{\text{2D}}^{2}-16t^{2}}}+\frac{K\left(\frac{16t^{2}}{E_{\text{2D}}^{2}}\right)}{E_{\text{2D}}}=\frac{\pi}{\varepsilon_{0}},\qquad\left|E_{\text{2D}}\right|<4t.\label{eq:2D-Spectrum}
\end{equation}
where $K\left(x\right)$ denotes an elliptic integral of the first
kind. The eigenenergy of the bound state can be calculated by solving
the Eq.~(\ref{eq:2D-Spectrum}) numerically for the energy $E_{\text{2D}}$. The corresponding eigenstates of the single-impurity system have the form,
\begin{widetext}
\begin{equation}
\psi(x,y)=\frac{\text{\ensuremath{\varepsilon_{0}\psi}(0,0)}}{2\pi}\intop_{-\pi}^{\pi}dk_{x}\frac{e^{ik_{x}x}}{E_{\text{2D}}-4t\sin^{2}\frac{k_{x}}{2}}{}_{3}\tilde{F}_{2}(\frac{1}{2},1,1;1-y,1+y;-\frac{4t}{E_{\text{2D}}-4t\sin^{2}\frac{k_{x}}{2}}).\label{eq:Integral2D}
\end{equation}
\end{widetext}
This integral (Eq. (\ref{eq:Integral2D})) can be numerically solved for the eigenstates of the 2D impurity system as illustrated in Fig.~\ref{fig:2D-Eigenstates}. One can clearly see that the decay of the eigenstate distributions of the system under the influence of single-site impurity decays exponentially as expected for the cases when (a) $\varepsilon_{0}=0.5,$ (b) $\varepsilon_{0}=1.5,$ and (c) $\varepsilon_{0}=3$, corresponding to the extended, critical, and localized regimes of the single-site impurity bound state, respectively. The eigenstate distributions decay symmetrically with site distance away from the impurity site. It is important to mention that introducing non-Hermiticity in the system may induce asymmetrical distributions of the eigenstates. In the extended regime, the impurity-induced state is distributed over all lattice sites, whereas it is distributed over a few lattice sites in the localized regime, as shown in Fig.~\ref{fig:2D-Eigenstates} (a) and Fig.~\ref{fig:2D-Eigenstates} (c), respectively. This shows that the eigenstate distributions may also characterize the localization-delocalization transition of the bound state.
\subsection{The 3D Impurity Lattice}
Turning to the case where a single impurity is embedded in a 3D noninteracting lattice model by keeping PBCs. The eigenstates of the 3D system Hamiltonian $\mathcal{H}_{\text{3D}}$ are plane waves in the absence of impurity with corresponding eigenenergy, $E_{\text{3D}}=-2t(\cos k_{x}+\cos k_{y}+ \cos k_{z})$. All eigenstates of the system Hamiltonian are translationally invariant, reflecting the metallic nature of the system. By introducing a single-site impurity, the translational invariance is broken and the impurity-induced state undergoes a localization-delocalization transition at a critical single-impurity potential, as illustrated in Fig.~\ref{fig:3D-ipr}. Importantly, the IPR approaches to zero as $L^{\alpha}$, with $\alpha<0$, in the extended regime (impurity strength $\varepsilon_{0}\le4.1$). On the other hand, the IPR increases as $L^{\alpha}$ with $\alpha\approx0$, in the critical regime for $4.1 \lessapprox \varepsilon_{0} \lessapprox 4.6$. The impurity state displays size-independent localized behavior for $\varepsilon_{0} \gtrapprox 4.6$, with IPR $\approx 1$.

The eigenstates distributions in the presence of the single-impurity potential of the 3D Hamiltonian can be obtained through the Green\textquoteright s
Functions:
\begin{equation}
\psi(x,y,z)=\frac{\varepsilon_{0}\delta_{x,0}\delta_{y,0}\delta_{z,0}}{E_{\text{3D}}-\mathcal{H}_{\text{3D}}}\psi(0,0,0),\label{eq:1D-GreenFtn-1-1}
\end{equation}
The quasimomenta corresponding to the eigenstates $\psi(x,y,z)$ can
be obtained by making a Fourier transformation of $\psi(x,y,z)$, i.e.,
\[\psi(k_{x},k_{y},k_{z})=\intop_{\Omega}dxdydz\psi(x,y,z)e^{-i(k_{x}x+k_{y}y+k_{z}z)}.
\]
The integral gives the following simplified expression:
\begin{align}
\psi(k_{x},k_{y},k_{z}) & =\frac{\varepsilon_{0}\psi(0,0,0)}{E_{\text{3D}}-2t(\cos k_{x}+\cos k_{y}+\cos k_{z})},\label{eq:-2-1}
\end{align}
Transforming back to the position space by making the inverse Fourier transformation of $\psi(k_{x},k_{y},k_{z})$ gives \cite{Inoue1974},
\begin{figure}[ht!]
\begin{centering}
\includegraphics[scale=0.36]{3D-State.eps}
\par\end{centering}
\caption{(Color online) The spatial profile of the} impurity induced quantum state in the 3D cubic lattice model.The eigenstate distribution is symmetrical around $x=0$ and is exponentially localized.
\label{fig:3D-State}
\end{figure}
%
\begin{align}
\psi(x,y,z) & =\varepsilon_{0}\psi(0,0,0)\mathcal{I}
.\label{eq:3D-State}
\end{align}
with
\begin{align}
\mathcal{I} & =\frac{1}{(2\pi)^{3}}\intop_{-\pi}^{\pi}\intop_{-\pi}^{\pi}\intop_{-\pi}^{\pi}\frac{e^{i(k_{x}x+k_{y}y+k_{z}z)}dk_{x}dk_{y}dk_{z}}{E_{\text{3D}}-2t(\cos k_{x}+\cos k_{y}+\cos k_{z})}
.\label{eq:3D-State0}
\end{align}
interpreted as the Fourier transform of the lattice Green's function for a cubic lattice. For $x=y=z=0$, the integral reduces to the density of states or the local Green's function at the origin. In order to find the eigenenergy of the bound state, we need to solve the expression~(\ref{eq:3D-State}) at the impurity embedded site $(x=y=z=0)$. The expression becomes:
\begin{align}
\mathcal{I} & =\frac{1}{\varepsilon_{0}}\label{eq:int}
\end{align}
For $|E_{3D}|<6$, the integral diverges (due to poles in the denominator), so $\mathcal{I}$ is not finite and Eq.~(\ref{eq:int}) has no solution. Similarly, at the band edge ($|E_{3D}|=6$) the integral diverges logarithmically. Thus, no finite solution exists at the band edges. For $|E_{3D}|> 6$, the Green's function is real and decreases monotonically with $E_{3D}$, and the integral can be expressed in terms of complete elliptic integral of the first kind,
\begin{align}
\frac{1}{\sqrt{E_{3D}^2-12}}\textbf{K}(\frac{4}{E_{3D}+\sqrt{E_{3D}^2-12}}) & =\frac{1}{\varepsilon_{0}}\label{eq:solution}
\end{align}
where $\textbf{K}(q)$ is the complete elliptical function of the $1^{\text{st}}$ kind defined as \citep{Morita1971-Elliptic}
\begin{align}
\textbf{K}(q) & = \intop_{0}^{\pi/2}\frac{d\theta}{\sqrt{1- q^{2} \sin^{2}{\theta}}}\nonumber\\
& = \frac{\pi}{2} {}_{2}F_{1}(\frac{1}{2},\frac{1}{2},1;q^{2}),\label{eq:Elliptical}
\end{align}
with
\begin{align}
q & = \frac{4}{E_{3D}+\sqrt{E_{3D}^2-12}}.
\end{align}
The expression [Eq. (\ref{eq:Elliptical})] can also be written as,
\begin{align}
\textbf{K}(q) & = \frac{\pi}{2} \sum _{n=0}^{\infty} \left( \frac{(\frac{1}{2})_{n}}{n!}\right)^{2} q^{2n},\label{eq:Elliptical-Sum}
\end{align}
by using the following identity
\begin{align}
{}_{2}F_{1}(\alpha ,\beta ;\gamma ;\chi)=\sum _{n=0}^{\infty} \frac{\Gamma(n+\alpha) \Gamma(n+\beta)  \Gamma(\gamma)}{n! \Gamma(\alpha)\Gamma(\beta) \Gamma(n+\gamma)} \chi^{n},
\end{align}
where $\Gamma(s)$ is the Gamma function and $(\frac{1}{2})_{n}=\Gamma(n+\frac{1}{2})/\Gamma(\frac{1}{2})$.
The expression given in Eq.~(\ref{eq:solution}) can be solved numerically for the calculations of the eigenenergy $E_{\text{3D}}$ of the bound state. In order to obtain the eigenstates of the single-impurity system, we first express the denominator of Eq.~(\ref{eq:3D-State0}) as a Laplace transform. The corresponding equation in an auxiliary integral representation is,
\begin{widetext}
\begin{align}
\psi(x,y,z) &= \frac{\varepsilon_{0}\psi(0,0,0)}{(2\pi)^{3}}\intop_{0}^{\infty}\intop_{-\pi}^{\pi}\intop_{-\pi}^{\pi}\intop_{-\pi}^{\pi}e^{i(k_{x}x+k_{y}y+k_{z}z)}e^{-\tau(E_{\text{3D}}-2(\cos k_{x}+\cos k_{y}+\cos k_{z}))} dk_{x}dk_{y}dk_{z} d\tau,\label{eq:Laplace}
\end{align}
\end{widetext}
The identity we are referring to is:
\begin{align}
\frac{1}{\beta} &= \intop_{0}^{\infty}e^{-\tau\beta} d\tau, \quad \text{Re}[\beta]>0.\label{eq:Identity}
\end{align}
The expression [Eq.~(\ref{eq:Laplace})] can be further simplified by expressing it in terms of the modified Bessel functions of the first kind, as given by (similar derivation can be found in Ref.\cite{Delves2007,Joyce1973}),
\begin{align}
\psi(x,y,z) &= \varepsilon_{0}\psi(0,0,0)\intop_{0}^{\infty} e^{-\tau E_{\text{3D}}} I_{x}(2\tau) I_{y}(2\tau)I_{z}(2\tau) d\tau,\label{eq:3Dsttate}
\end{align}
where
\begin{align}
I_{s}(2\tau) &=\frac{1}{2\pi} \intop_{-\pi}^{\pi}e^{ik_{s}s+2 \tau \cos k_{s}}dk_{s}; \quad s\in(x,y,z),
\label{eq:bessel}
\end{align}
is the modified Bessel functions of the first kind. The eigenstates of the 3D system can be obtained numerically by solving the expression given in Eq.~(\ref{eq:3Dsttate}). Crucially, the spatial distribution of eigenstates in a lattice system provides critical insights into the nature of impurity state-whether it is extended (metallic) or localized (insulating). The introduction of a single impurity into an otherwise clean lattice alters the system's behavior, inducing a spatial localized bound state and breaking translational symmetry. The distributions of the bound state is shown in Fig.~\ref{fig:3D-State}. The eigenstates exhibit mirror-symmetric spatial decay around the impurity site $x=0$, reflecting the isotropy of the lattice.
\section{Conclusions and Outlook}
In this work, we investigated impurity-induced localization in non-interacting lattice systems across one, two, and three spatial dimensions under periodic boundary conditions. By analyzing the inverse participation ratio and the spatial structure of eigenstates, we demonstrated that a single on-site impurity can induce a sharp crossover in the spatial character of impurity-bound states, which evolve from extended to strongly localized behavior beyond a critical impurity strength. While in one dimension this localization occurs readily, in higher dimensions the transition requires a larger impurity potential, reflecting the increased connectivity of the lattice; importantly, in all cases the bulk eigenstates remain extended, confirming that the observed transition is not a global Anderson localization. These results establish impurity-induced localization as a robust and dimension-dependent phenomenon and provide a unified framework for understanding how local perturbations shape eigenstate properties in lattice systems. Our findings open avenues for future studies involving interactions, multiple impurities, and dynamical effects in higher-dimensional settings.
\appendix
\bibliographystyle{apsrev4-2.bst}
\bibliography{Impurity}

@article{Delves2007,
  title={Derivation of exact product forms for the simple cubic lattice Green function using Fourier generating functions and Lie group identities},
  author={Delves, RT and Joyce, GS},
  journal={J. Phys. A: Math. Theor.},
  volume={40},
  number={29},
  pages={8329},
  year={2007},
  publisher={IOP Publishing}
}

@article{Joyce1973,
    author = {Joyce, G. S.},
    title = {On the simple cubic lattice Green function},
    journal = {Philos. Trans. A Math. Phys. Eng. Sci.},
    volume = {273},
    number = {1236},
    pages = {583-610},
    year = {1973},
    month = {02},
    issn = {0080-4614},
    doi = {10.1098/rsta.1973.0018},
    url = {https://doi.org/10.1098/rsta.1973.0018}
}

@article{greiner2002quantum,
  title={Quantum phase transition from a superfluid to a Mott insulator in a gas of ultracold atoms},
  author={Greiner, Markus and Mandel, Olaf and Esslinger, Tilman and H{\"a}nsch, Theodor W and Bloch, Immanuel},
  journal={nature},
  volume={415},
  number={6867},
  pages={39--44},
  year={2002},
  publisher={Nature Publishing Group UK London},
 url={ https://doi.org/10.1038/415039a},
doi={10.1038/415039a}
}

@article{Ye2025,
  title = {Disentangling connection between static and dynamical phase transitions},
  author = {Ye, Shihao and Khan, Niaz Ali and Sajid, Muhammad},
  journal = {Phys. Rev. A},
  volume = {111},
  issue = {4},
  pages = {042208},
  numpages = {9},
  year = {2025},
  month = {Apr},
  publisher = {American Physical Society},
  doi = {10.1103/PhysRevA.111.042208},
  url = {https://link.aps.org/doi/10.1103/PhysRevA.111.042208}
}

@article{Davies2025,
  title = {Two-scale effective model for defect-induced localization transitions in non-Hermitian systems},
  author = {Davies, B. and Barandun, S. and Hiltunen, E. O. and Craster, R. V. and Ammari, H.},
  journal = {Phys. Rev. B},
  volume = {111},
  issue = {3},
  pages = {035109},
  numpages = {10},
  year = {2025},
  month = {Jan},
  publisher = {American Physical Society},
  doi = {10.1103/PhysRevB.111.035109},
  url = {https://link.aps.org/doi/10.1103/PhysRevB.111.035109}
}

@book{Economou2006,
  title={Green's Functions in Quantum Physics},
  author={Economou, E.N.},
  isbn={9783540288411},
  lccn={2006926231},
  series={Springer Series in Solid-State Sciences},
  url={https://books.google.com/books?id=HdJDAAAAQBAJ},
  year={2006},
  publisher={Springer Berlin Heidelberg}
}

@article{Niaz2024,
  title = {Chebyshev polynomial approach to Loschmidt echo: Application to quench dynamics in two-dimensional quasicrystals},
  author = {Khan, Niaz Ali and Ye, Shihao and Zhou, Ziheng and Cheng, Shujie and Xianlong, Gao},
  journal = {Phys. Rev. E},
  volume = {109},
  issue = {6},
  pages = {065311},
  numpages = {9},
  year = {2024},
  month = {Jun},
  publisher = {American Physical Society},
  doi = {10.1103/PhysRevE.109.065311},
  url = {https://link.aps.org/doi/10.1103/PhysRevE.109.065311}
}

@article{KHAN2024114975,
title = {Fidelity susceptibility probes of dynamical quantum criticality},
journal = {Chaos, Solitons \& Fractals},
volume = {183},
pages = {114975},
year = {2024},
issn = {0960-0779},
doi = {https://doi.org/10.1016/j.chaos.2024.114975},
url = {https://www.sciencedirect.com/science/article/pii/S0960077924005277},
author = {Niaz Ali Khan},
}

@article{Kogan2024,
title = {Green’s Functions and DOS for Some 2D Lattices},
journal = {Graphene},
volume = {10},
pages = {1-12},
year = {2024},
issn = {0960-0779},
doi = {doi: 10.4236/graphene.2021.101001},
url = {https://www.scirp.org/journal/paperinformation?paperid=104591},
author = {Eugene Kogan, Godfrey Gumbs},
}

@article{Ye2024,
  title = {Energy-dependent dynamical quantum phase transitions in quasicrystals},
  author = {Ye, Shihao and Zhou, Ziheng and Khan, Niaz Ali and Xianlong, Gao},
  journal = {Phys. Rev. A},
  volume = {109},
  issue = {4},
  pages = {043319},
  numpages = {11},
  year = {2024},
  month = {Apr},
  publisher = {American Physical Society},
  doi = {10.1103/PhysRevA.109.043319},
  url = {https://link.aps.org/doi/10.1103/PhysRevA.109.043319}
}

@Article{Niaz2023,
  author   = {Khan, Niaz Ali and Wang, Pei and Jan, Munsif and Xianlong, Gao},
  journal  = {Scientific Reports},
  title    = {Anomalous correlation-induced dynamical phase transitions},
  year     = {2023},
  issn     = {2045-2322},
  number   = {1},
  pages    = {9470},
  volume   = {13},
  doi      = {10.1038/s41598-023-36564-9},
  refid    = {Khan2023},
  url      = {https://doi.org/10.1038/s41598-023-36564-9},
}

@article{bloch2022superfluid,
  title={The superfluid-to-Mott insulator transition and the birth of experimental quantum simulation},
  author={Bloch, Immanuel and Greiner, Markus},
  journal={Nat. Rev. Phys.},
  volume={4},
  number={12},
  pages={739--740},
  year={2022},
  publisher={Nature Publishing Group UK London},
 url={ https://doi.org/10.1038/s42254-022-00520-9},
doi={ 10.1038/s42254-022-00520-9}
}

@article{cao2022,
  title = {Superfluid--Mott-insulator quantum phase transition in a cavity optomagnonic system},
  author = {Cao, Qian and Tan, Lei and Liu, Wu-Ming},
  journal = {Phys. Rev. A},
  volume = {105},
  issue = {4},
  pages = {043705},
  numpages = {10},
  year = {2022},
  month = {Apr},
  publisher = {American Physical Society},
  doi = {10.1103/PhysRevA.105.043705},
  url = {https://link.aps.org/doi/10.1103/PhysRevA.105.043705}
}

@article{PhysRevLett.126.116401,
  title = {Quantum Phase Transition in a Quantum Ising Chain at Nonzero Temperatures},
  author = {Zhang, K. L. and Song, Z.},
  journal = {Phys. Rev. Lett.},
  volume = {126},
  issue = {11},
  pages = {116401},
  numpages = {6},
  year = {2021},
  month = {Mar},
  publisher = {American Physical Society},
  doi = {10.1103/PhysRevLett.126.116401},
  url = {https://link.aps.org/doi/10.1103/PhysRevLett.126.116401}
}

@article{PhysRevB.81.205102,
  title = {Quantum phase transition from an antiferromagnet to a spin liquid in a metal},
  author = {Grover, Tarun and Senthil, T.},
  journal = {Phys. Rev. B},
  volume = {81},
  issue = {20},
  pages = {205102},
  numpages = {8},
  year = {2010},
  month = {May},
  publisher = {American Physical Society},
  doi = {10.1103/PhysRevB.81.205102},
  url = {https://link.aps.org/doi/10.1103/PhysRevB.81.205102}
}

@article{paschen2021quantum,
  title={Quantum phases driven by strong correlations},
  author={Paschen, Silke and Si, Qimiao},
  journal={Nat. Rev. Phys.},
  volume={3},
  number={1},
  pages={9--26},
  year={2021},
  publisher={Nature Publishing Group UK London},
 url={ https://doi.org/10.1038/s42254-020-00262-6},
doi={10.1038/s42254-020-00262-6}
}

@article{zheng2022quantum,
  title={Quantum criticality of excitonic Mott metal-insulator transitions in black phosphorus},
  author={Zheng, Binjie and Wang, Junzhuan and Wang, Qianghua and Su, Xin and Huang, Tianye and Li, Songlin and Wang, Fengqiu and Shi, Yi and Wang, Xiaomu},
  journal={Nat. Commun.},
  volume={13},
  number={1},
  pages={7797},
  year={2022},
  publisher={Nature Publishing Group UK London},
url={ https://doi.org/10.1038/s41467-022-35567-w},
doi={10.1038/s41467-022-35567-w}
}

@Article{Anderson1958,
  author    = {Anderson, P. W.},
  title     = {Absence of Diffusion in Certain Random Lattices},
  journal   = {Phys. Rev.},
  year      = {1958},
  volume    = {109},
  pages     = {1492--1505},
  month     = {Mar},
  doi       = {10.1103/PhysRev.109.1492},
  issue     = {5},
  numpages  = {0},
  publisher = {American Physical Society},
  url       = {https://link.aps.org/doi/10.1103/PhysRev.109.1492},
}

@article{PhysRevB.95.144204,
  title = {Localization landscape theory of disorder in semiconductors. I. Theory and modeling},
  author = {Filoche, Marcel and Piccardo, Marco and Wu, Yuh-Renn and Li, Chi-Kang and Weisbuch, Claude and Mayboroda, Svitlana},
  journal = {Phys. Rev. B},
  volume = {95},
  issue = {14},
  pages = {144204},
  numpages = {18},
  year = {2017},
  month = {Apr},
  publisher = {American Physical Society},
  doi = {10.1103/PhysRevB.95.144204},
  url = {https://link.aps.org/doi/10.1103/PhysRevB.95.144204}
}

@article{poduval2023anderson,
  title = {Anderson localization in doped semiconductors},
  author = {Poduval, Prathyush P. and Das Sarma, Sankar},
  journal = {Phys. Rev. B},
  volume = {107},
  issue = {17},
  pages = {174204},
  numpages = {14},
  year = {2023},
  month = {May},
  publisher = {American Physical Society},
  doi = {10.1103/PhysRevB.107.174204},
  url = {https://link.aps.org/doi/10.1103/PhysRevB.107.174204}
}

@ARTICLE{Mafi2019,
  author={Mafi, Arash and Ballato, John and Koch, Karl W. and Schülzgen, Axel},
  journal={Journal of Lightwave Technology}, 
  title={Disordered Anderson Localization Optical Fibers for Image Transport—A Review}, 
  year={2019},
  volume={37},
  number={22},
  pages={5652-5659},
  doi={10.1109/JLT.2019.2916020}}

@article{white2020observation,
  title={Observation of two-dimensional Anderson localisation of ultracold atoms},
  author={White, Donald H and Haase, Thomas A and Brown, Dylan J and Hoogerland, Maarten D and Najafabadi, Mojdeh S and Helm, John L and Gies, Christopher and Schumayer, Daniel and Hutchinson, David AW},
  journal={Nat. Commu.},
  volume={11},
  number={1},
  pages={4942},
  year={2020},
  publisher={Nature Publishing Group UK London},
url={ https://doi.org/10.1038/s41467-020-18652-w},
doi={ 10.1038/s41467-020-18652-w}
}

@article{AGNE20212970,
title = {Disorder-induced Anderson-like localization for bidimensional thermoelectrics optimization},
journal = {Matter},
volume = {4},
number = {9},
pages = {2970-2984},
year = {2021},
issn = {2590-2385},
doi = {https://doi.org/10.1016/j.matt.2021.07.017},
url = {https://www.sciencedirect.com/science/article/pii/S2590238521003647},
author = {Matthias T. Agne and Felix R.L. Lange and James P. Male and K. Simon Siegert and Hanno Volker and Christian Poltorak and Annika Poitz and Theo Siegrist and Stefan Maier and G. Jeffrey Snyder and Matthias Wuttig}
}

@article{jendrzejewski2012three,
  title={Three-dimensional localization of ultracold atoms in an optical disordered potential},
  author={Jendrzejewski, Fred and Bernard, Alain and Mueller, Killian and Cheinet, Patrick and Josse, Vincent and Piraud, Marie and Pezz{\'e}, Luca and Sanchez-Palencia, Laurent and Aspect, Alain and Bouyer, Philippe},
  journal={Nat. Phys.},
  volume={8},
  number={5},
  pages={398--403},
  year={2012},
  publisher={Nature Publishing Group UK London},
 url={ https://doi.org/10.1038/nphys2256},
doi={ 10.1038/nphys2256}
}

@Book{Thouless1979,
  title     = {Ill-Condensed Matter, Les Houches Session XXXI},
  publisher = {North-Holland, New York},
  year      = {1979},
  author    = {D. J. Thouless},
  editor    = {R. Balian, R. Maynard and G. Toulouse},
  url       = {https://www.amazon.sg/Ill-condensed-Matter-Houches-Session-Xxxi/dp/9971950596},
}

@Article{Niaz2020,
  author  = {N. A. Khan and J. P. S. Pires and J. M. V. P. Lopes and J. M. B. L. dos Santos},
  title   = {Probing the Global Delocalization Transition in the de Moura-Lyra Model with the Kernel Polynomial Method},
  journal = {EPJ Web Conf.},
  year    = {2020},
  volume  = {233},
  pages   = {05011},
  doi     = {10.1051/epjconf/202023305011},
}

@Article{Niaz2019,
  author    = {N. A. Khan and J. M. V. P. Lopes and J. P. S. Pires and J. M. B. L. dos Santos},
  title     = {Spectral functions of one-dimensional systems with correlated disorder},
  journal   = {J. Phys.: Condens. Matter},
  year      = {2019},
  volume    = {31},
  number    = {17},
  pages     = {175501},
  month     = {feb},
  doi       = {10.1088/1361-648x/ab03ad},
  publisher = {{IOP} Publishing},
}

@Article{Pires2019,
  author    = {Pires, J. P. S. and Khan, N. A. and Lopes, J. M. V. P. and dos Santos, J. M. B. L.},
  title     = {Global delocalization transition in the de Moura--Lyra model},
  journal   = {Phys. Rev. B},
  year      = {2019},
  volume    = {99},
  pages     = {205148},
  month     = {May},
  doi       = {10.1103/PhysRevB.99.205148},
  issue     = {20},
  numpages  = {6},
  publisher = {American Physical Society},
}

@Article{Niaz2022,
  author   = {N. A. Khan and S. Muhammad and M. Sajid},
  title    = {Single parameter scaling in the correlated Anderson model},
  journal  = {Physica E},
  year     = {2022},
  volume   = {139},
  pages    = {115150},
  issn     = {1386-9477},
  doi      = {https://doi.org/10.1016/j.physe.2022.115150},
  url      = {https://www.sciencedirect.com/science/article/pii/S1386947722000169},
}

@Article{Niaz2021,
  author    = {N. A. Khan and S. T. Amin},
  title     = {Probing band-center anomaly with the Kernel polynomial method},
  journal   = {Phys. Scr.},
  year      = {2021},
  volume    = {96},
  number    = {4},
  pages     = {045812},
  month     = {Feb},
  doi       = {10.1088/1402-4896/abe322},
  publisher = {{IOP} Publishing},
  url       = {https://doi.org/10.1088/1402-4896/abe322},
}

@Inbook{Fabrizio2022,
author="Fabrizio, Michele",
title="Kondo Effect and the Physics of the Anderson Impurity Model",
year="2022",
publisher="Springer International Publishing",
address="Cham",
isbn="978-3-031-16305-0",
doi="10.1007/978-3-031-16305-0_7",
url="https://doi.org/10.1007/978-3-031-16305-0_7"
}

@article{PhysRevB.107.125103,
  title = {Real-time evolution of Anderson impurity models via tensor network influence functionals},
  author = {Ng, Nathan and Park, Gunhee and Millis, Andrew J. and Chan, Garnet Kin-Lic and Reichman, David R.},
  journal = {Phys. Rev. B},
  volume = {107},
  issue = {12},
  pages = {125103},
  numpages = {8},
  year = {2023},
  month = {Mar},
  publisher = {American Physical Society},
  doi = {10.1103/PhysRevB.107.125103},
  url = {https://link.aps.org/doi/10.1103/PhysRevB.107.125103}
}

@Article{Lagendijk2009,
  author  = {Lagendijk,A. and Tiggelen,B. V. and Wiersma,D. S.},
  title   = {Fifty years of Anderson localization},
  journal = {Phys. Today},
  year    = {2009},
  volume  = {62},
  number  = {8},
  pages   = {24-29},
  doi     = {10.1063/1.3206091},
  url     = { 
        https://doi.org/10.1063/1.3206091
    
},
}

@Article{Roati2008,
  author   = {Roati, G. and D`Errico, C. and Fallani, L. and Fattori, M. and Fort, C. and Zaccanti, M. and Modugno, G. and Modugno, M. and Inguscio, M.},
  title    = {Anderson localization of a non-interacting Bose-Einstein condensate},
  journal  = {Nature},
  year     = {2008},
  volume   = {453},
  number   = {7197},
  pages    = {895--898},
  month    = jun,
  issn     = {1476-4687},
  refid    = {Roati2008},
  url      = {https://doi.org/10.1038/nature07071},
}

@Book{Muller2010,
  title     = {Disorder and interference: localization phenomena: Ultracold Gases and Quantum Information,},
  publisher = {(Les Houches Summer School Session XCI ) ed C Miniatura et al (Oxford: Oxford University Press)},
  year      = {2010},
  author    = {C. A. Muller and D. Delande},
}

@Article{Niaz2022-EC1DCD,
  author   = {Niaz Ali Khan and Munsif Jan and Gao Xianlong},
  title    = {Entanglement contour in the disordered electronic systems},
  journal  = {Physica E},
  year     = {2023},
  volume   = {145},
  pages    = {115511},
  issn     = {1386-9477},
  doi      = {https://doi.org/10.1016/j.physe.2022.115511},
  url      = {https://www.sciencedirect.com/science/article/pii/S1386947722003344},
}

@article{Molignini2023,
  title = {Anomalous skin effects in disordered systems with a single non-Hermitian impurity},
  author = {Molignini, Paolo and Arandes, Oscar and Bergholtz, Emil J.},
  journal = {Phys. Rev. Res.},
  volume = {5},
  issue = {3},
  pages = {033058},
  numpages = {10},
  year = {2023},
  month = {Jul},
  publisher = {American Physical Society},
  doi = {10.1103/PhysRevResearch.5.033058},
  url = {https://link.aps.org/doi/10.1103/PhysRevResearch.5.033058}
}

@article{Wang2025,
  title = {Multiple physical quantities sensors based on non-Hermitian topological systems with an impurity},
  author = {Wang, J. J. and Zhang, Yubin and Zhang, Xiaomin and Yi, X. X.},
  journal = {Phys. Rev. Res.},
  volume = {7},
  issue = {3},
  pages = {033133},
  numpages = {11},
  year = {2025},
  month = {Aug},
  publisher = {American Physical Society},
  doi = {10.1103/h9b6-md2f},
  url = {https://link.aps.org/doi/10.1103/h9b6-md2f}
}

@article{Hetenyi2025,
  title = {Localized states and skin effect around non-Hermitian impurities in tight-binding models},
  author = {Het\'enyi, Bal\'azs and D\'ora, Bal\'azs},
  journal = {Phys. Rev. B},
  volume = {112},
  issue = {7},
  pages = {075123},
  numpages = {12},
  year = {2025},
  month = {Aug},
  publisher = {American Physical Society},
  doi = {10.1103/xbj1-hfyf},
  url = {https://link.aps.org/doi/10.1103/xbj1-hfyf}
}

@article{Leumer2025,
    author = {Leumer, Nico G. and Bercioux, Dario},
    title = {Impurity-induced counter skin-effect and linear modes in the Hatano–Nelson model},
    journal = {APL Quantum},
    volume = {2},
    number = {3},
    pages = {036105},
    year = {2025},
    month = {09},
    issn = {2835-0103},
    doi = {10.1063/5.0270893},
    url = {https://doi.org/10.1063/5.0270893}
}

@article{Roccati2021,
  title = {Non-Hermitian skin effect as an impurity problem},
  author = {Roccati, Federico},
  journal = {Phys. Rev. A},
  volume = {104},
  issue = {2},
  pages = {022215},
  numpages = {7},
  year = {2021},
  month = {Aug},
  publisher = {American Physical Society},
  doi = {10.1103/PhysRevA.104.022215},
  url = {https://link.aps.org/doi/10.1103/PhysRevA.104.022215}
}

@article{Inoue1974,
    author = {Inoue, Michiko},
    title = {Lattice Green's function for the face centered cubic lattice},
    journal = {J. Math. Phys.},
    volume = {15},
    number = {6},
    pages = {704-707},
    year = {1974},
    month = {06},
    issn = {0022-2488},
    doi = {10.1063/1.1666714},
    url = {https://doi.org/10.1063/1.1666714},
}

@Article{Liu2020,
  author    = {Liu, Yanxia and Chen, Shu},
  title     = {Diagnosis of bulk phase diagram of nonreciprocal topological lattices by impurity modes},
  journal   = {Phys. Rev. B},
  year      = {2020},
  volume    = {102},
  pages     = {075404},
  month     = {Aug},
  doi       = {10.1103/PhysRevB.102.075404},
  issue     = {7},
  numpages  = {10},
  publisher = {American Physical Society},
  url       = {https://link.aps.org/doi/10.1103/PhysRevB.102.075404},
}

@Article{Liu2021,
  author    = {Liu, Yanxia and Zeng, Yumeng and Li, Linhu and Chen, Shu},
  title     = {Exact solution of the single impurity problem in nonreciprocal lattices: Impurity-induced size-dependent non-Hermitian skin effect},
  journal   = {Phys. Rev. B},
  year      = {2021},
  volume    = {104},
  pages     = {085401},
  month     = {Aug},
  doi       = {10.1103/PhysRevB.104.085401},
  issue     = {8},
  numpages  = {9},
  publisher = {American Physical Society},
  url       = {https://link.aps.org/doi/10.1103/PhysRevB.104.085401},
}

@article{Morita1971-Elliptic,
author = {Morita,Tohru  and Horiguchi,Tsuyoshi },
title = {Lattice Green's Functions for the Cubic Lattices in Terms of the Complete Elliptic Integral},
journal = {J. Math. Phys.},
volume = {12},
number = {6},
pages = {981-986},
year = {1971},
doi = {10.1063/1.1665692},
URL = {https://doi.org/10.1063/1.1665692},
}

\end{document}